\documentclass[10pt,letterpaper]{article}
\usepackage[top=0.85in,left=2.75in,footskip=0.75in]{geometry}

\usepackage{amsmath,amssymb}

\usepackage{changepage}

\usepackage[utf8x]{inputenc}

\usepackage{textcomp,marvosym}

\usepackage{cite}

\usepackage{nameref,hyperref}

\usepackage[right]{lineno}

\usepackage{microtype}
\DisableLigatures[f]{encoding = *, family = * }

\usepackage[table]{xcolor}

\usepackage{array}

\usepackage{algorithm}
\usepackage{bm}% bold math
\usepackage{tabularx}
\usepackage{multicol}
\usepackage{booktabs}
\usepackage{threeparttable}
\usepackage{rotating}
\usepackage{dcolumn}% Align table columns on decimal point
\usepackage{appendix}
\usepackage{cleveref}
\usepackage[noend]{algpseudocode}
\usepackage{color}

\newcommand{\project}[1]{\textsl{#1}}
\newcommand{\entrofy}{\project{Entrofy}}
\DeclareMathOperator*{\argmax}{argmax}
\newcolumntype{+}{!{\vrule width 2pt}}

\newlength\savedwidth

\raggedright
\usepackage[aboveskip=1pt,labelfont=bf,labelsep=period,justification=raggedright,singlelinecheck=off]{caption}

\makeatletter
\renewcommand{\@biblabel}[1]{\quad#1.}
\makeatother

\usepackage{lastpage,fancyhdr,graphicx}
\usepackage{epstopdf}
\fancyheadoffset[L]{2.25in}
\fancyfootoffset[L]{2.25in}
\begin{document}
\vspace*{0.2in}

% Title must be 250 characters or less.
\begin{flushleft}
{\Large
\textbf\newline{Entrofy Your Cohort: A Data Science Approach to Candidate Selection} % Please use "sentence case" for title and headings (capitalize only the first word in a title (or heading), the first word in a subtitle (or subheading), and any proper nouns).
}
\newline
% Insert author names, affiliations and corresponding author email (do not include titles, positions, or degrees).
\\
Daniela Huppenkothen\textsuperscript{1\Yinyang\textcurrency},
Brian McFee\textsuperscript{2,3\Yinyang},
Laura Nor\'{e}n\textsuperscript{4\Yinyang}
\\
\bigskip
\textbf{1} DIRAC Institute, Department of Astronomy, University of Washington, 3910 15th Ave NE, Seattle, WA 98195 
\\
\textbf{2} Center for Data Science, New York University,  60 5th Ave., New York, NY 10003
\\
\textbf{3} Music and Audio Research Lab, New York University, 35 West 4th St., New York, NY 10003
\\
\textbf{4} Obsidian Security, 888 San Clemente Drive, Newport Beach, 92660
\\
\bigskip

% Insert additional author notes using the symbols described below. Insert symbol callouts after author names as necessary.
% 
% Remove or comment out the author notes below if they aren't used.
%
% Primary Equal Contribution Note
\Yinyang These authors contributed equally to this work.

% Additional Equal Contribution Note
% Also use this double-dagger symbol for special authorship notes, such as senior authorship.
%\ddag These authors also contributed equally to this work.

% Current address notes
%\textcurrency Current Address: Dept/Program/Center, Institution Name, City, State, Country % change symbol to "\textcurrency a" if more than one current address note
% \textcurrency b Insert second current address 
% \textcurrency c Insert third current address

% Deceased author note
%\dag Deceased

% Group/Consortium Author Note
%\textpilcrow Membership list can be found in the Acknowledgments section.

% Use the asterisk to denote corresponding authorship and provide email address in note below.
* dhuppenk@uw.edu

\end{flushleft}
% Please keep the abstract below 300 words
\section*{Abstract}
Selecting a cohort from a set of candidates is a common task within and beyond academia. Admitting students, awarding grants, choosing speakers for a conference are situations where human biases may affect the make-up of the final cohort. In this paper, we propose a new algorithm, Entrofy, designed to be part of a larger decision making strategy aimed at making cohort selection as just, quantitative, transparent, and accountable as possible.
We suggest this algorithm be embedded in a two-step selection procedure. First, all application materials are stripped of names, institutional affiliations, and other markers of identity that could induce conscious or sub-conscious bias. During blind review, the committee selects all applicants, submissions, or other entities that meet their merit-based criteria. This often yields a cohort larger than the admissible number.
In the second stage, the target cohort can be chosen from this meritorious pool via a new algorithm and software tool called Entrofy. Entrofy optimizes differences across an assignable set of categories selected by the human committee. Criteria could include gender, academic discipline, home country, experience with certain technologies, or other quantifiable characteristics. The Entrofy algorithm then yields the computational maximization of diversity by solving the tie-breaking problem with provable performance guarantees. We show how Entrofy selects cohorts according to pre-determined characteristics in simulated sets of applications and demonstrate its use in a case study of Astro Hack Week. 
This two stage cohort selection process allows human judgment to prevail when assessing merit, but assigns the assessment of diversity to a computational process less likely to be beset by human bias. Importantly, the stage at which diversity assessments occur is fully transparent and auditable with Entrofy. Splitting merit and diversity considerations into their own assessment stages makes it easier to explain why a given candidate was selected or rejected.

% Please keep the Author Summary between 150 and 200 words
% Use first person. PLOS ONE authors please skip this step. 
% Author Summary not valid for PLOS ONE submissions.   
%\section*{Author summary}
%One-sentence summary: We propose a new strategy to increase diversity, transparency, and accountability in cohort selection using a two-stage process beginning with a blinded merit review followed by the application of a newly proposed diversity-maximizing algorithm, Entrofy. \\

%\linenumbers

% Use "Eq" instead of "Equation" for Eqcitations.
\section*{\label{sec:introduction}Introduction}

Selecting a cohort from among a pool of applicants is a common task within and beyond academia. Common cohort selection scenarios include admitting students to degree programs, awarding fellowships and grants, selecting speakers for a seminar series, panel members for a conference, or participants for space-limited workshops. A frequent challenge in all of these cases is the inherent limitation of available slots or jobs, as well as the practical need to select the best cohort out of a much larger set of available candidates. At present, this type of selection in academia is generally performed by committees convened to judge each candidate or proposal, taking all of their personal and professional characteristics into account at once. 

We argue this process may yield suboptimal results due to the conflation of professional/academic merit and a range of other personal characteristics~\cite{bohnet2016}. With respect to professional and academic merit, the goal is generally to choose the best possible candidates. With respect to certain types of meritorious characteristics, say, the proven ability to publish, all candidates are competing with the same criteria in mind. With respect to other characteristics---for instance, strength in certain computational tasks, gender and racial background, or age range---it may be more advantageous to select a cohort that contains as much diversity as possible. We propose to improve on the typical selection process by introducing a computational tool that is aimed at assisting the process of assessing diversity across assignable criteria. Our assumptions are that: 1) merit review may be too nuanced for computational assessment and that 2) diversity maximization may be too computationally intensive and subject to human bias ~\cite{Greenwald1995} for committees to efficiently select optimal cohorts. Thus, we suggest to split the selection into a (possibly blinded) merit review process and an algorithmic diversity-maximization step where diversity criteria are defined by each cohort selection committee.

Much has been written about the inherent biases in selecting candidates for employment~\cite{reskin2000,riach2002,gorman2005,pager2005,krieger2006,rooth2010,kuncel2013}. In particular, despite ample evidence to the contrary~\cite{grove1996,grove2000}, is a prevailing opinion that successful candidate selection can be learned and intuitive judgments are predictive of future job performance. 
This often leads committee members to make overconfident predictions about the success of the selection procedure with respect to employee productivity, and an underestimation of the inherently stochastic processes involved~\cite{pulakos1996,kausel2016}. Additionally, humans tend to be swayed by stories over facts: in the employment context, this can lead to decisions that defy evidence and logic~\cite{highhouse1998}. Other inherent biases have been found to affect selection outcomes and salaries with respect to gender~\cite{mossracusin2012,reuben2014} and race~\cite{bertrand2004,lavergne2004} based on the candidate's name and/or appearance. In several replicated studies across different problems, unstructured interviews have been found to produce worse outcomes than grades, general intelligence tests, and structured interviews with clear selection criteria~\cite{kuncel2013,kausel2016}, yet selection procedures continue to be dominated by panels relying largely on experience, intuition, and debate over one applicant at a time~\cite{rivera2015} rather than systematic consideration of possible cohorts of similar collective merit. Additionally, panels tend to rely more heavily on the representativeness heuristic (assuming that because something is more representative, it is also more likely~\cite{kahnemann1972}) and show  more overconfidence in their judgments than individuals do~\cite{sunstein2015}.
These findings stand in contrast to the goal of most selection procedures: to choose the best set of candidates conditional on minimum requirements and goals dictated by the situation. 

While there is a vast literature regarding hiring practices, there is little empirical research about academic conference selection procedures. Still, conferences are an important cornerstone of academic careers: they provide venues for learning the most recent scientific results (often ahead of publication), for presenting one's own work, and for networking ~\cite{nicolson2016}. Conference presentations are considered an important measure of academic success~\cite{Buddeberg-Fischer2008} and increase research visibility. This is particularly the case for early-career researchers and those from international or lower-tier institutions~\cite{leite2014}. Workshop attendance~\cite{blau2010} and conference presentations~\cite{housri2008} can boost the probability of early-stage female faculty members getting published and cited. For example,~\cite{katerndahl1999} find that the number of conferences attended predicts post-conference publications, presentations, and current research activity for attendees of a research methods conference for primary care practitioners. 
There is thus a legitimate concern that the same biases affecting hiring practices may analogously influence selection of abstracts and participants at academic workshops and conferences. For example,~\cite{ross2006} find that for major medical conferences, an unblinded selection procedure favours candidates from the United States, English-speaking countries and from prestigious universities. 

The problem of selecting a diverse cohort from a candidate pool is related to diversity optimization within the context of search and information retrieval~\cite{carbonell1998use} where a set of documents are scored by relevance to a search query and a diverse set of results is selected to minimize information redundancy between elements of the set.
While ranking diversity bears many similarities to the problem of diverse cohort selection, there are two fundamental distinctions that render existing algorithmic solutions inapplicable to the present context.
First, typical ranking diversity methods conflate "relevance" scores (or \emph{merit} in our setting) with measures of diversity~\cite{clarke2008novelty}, which makes it impossible to determine or explain why any particular candidate was or was not selected.
Second, ranking diversity methods require a well-calibrated ordering of all candidates, which can be computationally (and politically) difficult to achieve when scores are produced by multiple judges~\cite{dwork2001rank}.
We opt for a clear separation between determination of individual merit and the creation of an optimally diverse cohort. Separating these two steps obviates the need for finding a consensus ranking of candidates and provides an audit trail for determining how each inclusion/exclusion decision was made. Accountability is critical for establishing trust in a system that directly impacts people~\cite{fatml}.
% diversity optimization is studied in IR

In this paper we introduce a new algorithm as part of a larger strategy to make cohort selection in academic and non-academic contexts more quantitative and transparent in order to allow committees to interrogate their own decisions, probe for inherent biases, and evaluate the efficacy of their decision making processes.   
We advocate for a two-step procedure including an initial blind selection for quality and a subsequent computer-assisted selection from the pool meritorious candidates that maximizes user-defined diversity measured at the cohort level. We suggest that the initial blind selection for quality should be made by individual committee members in isolation in order to minimize "groupthink" effects~\cite{janis1982}.
This step could, for example, incorporate a score given by each committee member to each applicant based on clearly articulated quality criteria.
Because blind selection procedures may lead to unexpected (potentially negative) consequences, these scores should themselves be evaluated for unwanted biases.
Any quality selection will likely be subject to a high intrinsic variance, much of which can be accounted for by heuristics and biases in human decision making ~\cite{Greenwald1995, grove1996}.

As explained above, we therefore suggest translating scores directly into a binary positive or negative decision about admissibility.
If the pool of admissible candidates exceeds the number of available spots, the initial quality selection is then followed in a second step by a computer-assisted selection of the cohort out of the quality-controlled pool conditioned on other constraints important to the success of the procedure (e.g., demographic diversity). We present a novel algorithm and accompanying software tool called Entrofy to assist the second stage of the process, framed as an optimization problem over a large number of competing variables. While the main paper lays out the algorithm and discusses its application using both simulations and a case study, the Supplementary Materials provide practical advice for using the algorithm and the associated software tool.

\section*{Methods}
\label{sec:algorithm}

\subsection*{Ethics Statement}
The experimental procedures were approved by the Institutional Review Board at New York University. The questionnaire results referenced in the Discussion section were obtained during a (voluntary) online post-workshop surveys in 2016 and 2017, respectively. Workshop participants gave consent online using an IRB-compliant form before beginning the survey. 

\subsection{Overview}

Given a pool of candidates, a collection of attributes describing each candidate, and target proportions of each attribute for the selected set, the goal is to find a subset of candidates whose statistics match the target proportions as closely as possible. 
For this purpose, we define an objective function that measures the relative improvement of adding a given candidate to the pool of selected participants, compared to the target selection criteria. Starting either from a group of pre-selected (e.g.\ invited) participants, or a random position, the algorithm runs sequentially through all candidates and computes the relative improvement of adding that candidate to the selected set compared to the target categories. The candidate with the highest improvement is added to the participant set, and the algorithm continues until the pre-defined number of participants is reached.

When two candidates would produce the same improvement, the tie is broken randomly.
In practice, for optimization over multiple categories this can lead to overall solutions that are suboptimal, but we show in Results that this problem can be effectively solved by picking the best solution out of several randomized runs of the algorithm. 

\subsection*{Algorithm}

Let $S$ denote the set of acceptable candidates.
Let ${a_i : S \rightarrow \{0, 1\}}$ denote the indicator function of the $i^\text{th}$ attribute.
Given a target set size $k \in \mathbb{N}$, and a set of target frequencies $p_i$ for each attribute, our
goal is to find a subset $X \subseteq S$ of size $|X|=k$ such that
\begin{equation}
    \forall i~:~\sum_{x \in X} a_i(x) \geq k p_i,\label{eq:entrofy}
\end{equation}
that is, the selected subset has statistics which meet or exceed the target proportions for each attribute $a_i$.
In full generality, there may not be any feasible solution if there are too few candidates possessing the desired attributes.
We therefore relax~\eqref{eq:entrofy} to a maximization of the following objective function:
\begin{eqnarray}
    f(X) &:=& \sum_i w_i f_i(X) \label{eq:entrofy-opt}\\
         &=& \sum_i w_i \min\left(k p_i, \sum_{x \in X} a_i(x)\right),\notag
\end{eqnarray}
where $w_i \geq 0$ are non-negative weights, and each $f_i$ measures the number of selected candidates $x$ that satisfy attribute $i$, but stops counting after $kp_i$ have been found.
The coefficients $w_i$ can be controlled by the user to express the relative importance among competing terms in the objective.

Both forms of the Entrofy problem are NP-Hard (Supplementary Materials), so we do not expect an efficient algorithm to produce exact solutions in all cases~\cite{gary1979computers}.
Instead, we will focus on developing an efficient, greedy approximation algorithm with provable guarantees.

The greedy maximization strategy (\Cref{alg:maximize}) operates by iteratively selecting the point $x$ with the maximum \emph{marginal gain} over the current solution $X\subseteq S$:
\begin{equation}
    \Delta f(X, x) := f\left(X \cup \{x\}\right) - f(X).
\end{equation}

\begin{algorithm}[H]
\caption{Greedy set-function maximization}\label{alg:maximize}
\begin{algorithmic}[1]
\Procedure{Maximize}{$f, S, k$}
\State{Initialize $X \leftarrow \emptyset$}
\While{$|X| < k$}
\State{$X \leftarrow X \cup \left\{\displaystyle\argmax_{x \in S \setminus X} \Delta f(X, x) \right\}$}
\EndWhile{}
\State{\textbf{return} $X$}
\EndProcedure{}
\end{algorithmic}
\end{algorithm}

If the objective function $f$ is \emph{monotone} (non-decreasing), non-negative, and \emph{submodular}, then a greedy maximization algorithm is guaranteed to find a solution $X^*$ in polynomial time such that $f(X^*) \geq (1 - e^{-1}) f^*$, where $f^*$ is the optimal solution value~\cite{fujishige2005submodular}.
We demonstrate in the Supplementary Materials
that~\eqref{eq:entrofy-opt} satisfies these conditions, and is therefore amenable to efficient, approximate optimization.

\subsection*{Concave transformation}
When optimizing selection over multiple attributes, a single element $x$'s contribution to the objective for each term $f_i$ is either 0 or 1, regardless of how far from the target proportion $kp_i$ the current solution lies.
Consequently, when attribute $i$ is close to being satisfied while another attribute $j$ is far from its target proportion, the greedy selection algorithm cannot distinguish between elements that improve $f_j$ or $f_i$.
As a result, the algorithm can produce poor solutions which reach the target frequencies for some attributes at the expense of others.

\begin{figure}
\begin{center}
\includegraphics[width=3.5in]{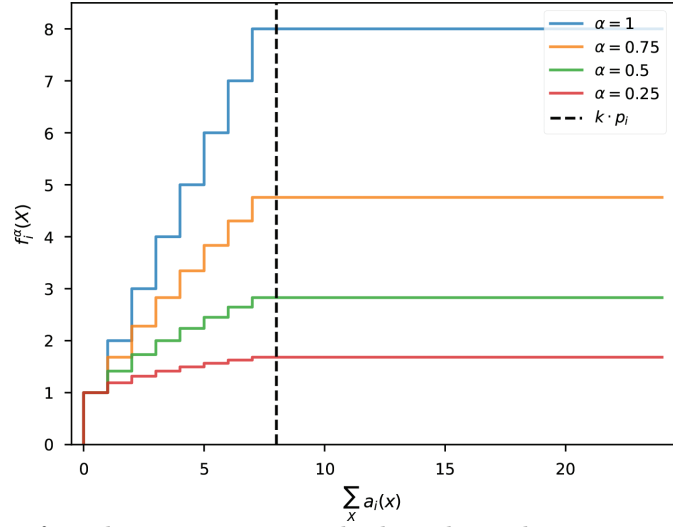}
\caption{Raising $f_i$ to the $0 < \alpha \leq 1$ power leads to diminishing marginal gains as the selected set approaches the target $kp_i$.}
\label{fig:concave}
\end{center}
\end{figure}

This problem can be avoided by applying a concave, monotone transformation to $f_i$, for example, $f_i \mapsto f_i^\alpha$ for some ${0 < \alpha \leq 1}$.
Under this transformation, the function remains non-negative, monotone, and concave, and is therefore still amenable to greedy maximization.
For values $\alpha < 1$, the marginal gain $\Delta f_i^\alpha(X, x)$ diminishes as $f_i$ increases, as illustrated in Fig \ref{fig:concave}.
The greedy selection algorithm is therefore more likely to select elements which improve coverage of attributes that are far from their target proportions.

\subsection*{Randomization and (near) tie-breaking}

The greedy maximization algorithm may have to select among multiple equally good options at each step.
Typically, ties are broken arbitrarily by selecting a maximizer at random.
When the objective includes multiple competing terms, this can introduce some variance in the objective value of the solutions produced by the algorithm.
This variance can be reduced by running the algorithm several times and selecting the result with highest objective value.

Additionally, the algorithm can be made more robust by allowing it to explore slightly sub-optimal local choices. We achieve this by relaxing the greedy maximization to select randomly among elements in the top $q$-quantile of marginal gain. For $q=1.0$, this reduces to the greedy maximization algorithm.

\Cref{alg:entrofy} lists the full Entrofy algorithm, including the concave transformation and randomized near-tie breaking.

\begin{algorithm}[H]
\caption{The full Entrofy algorithm}\label{alg:entrofy}
\begin{algorithmic}[1]
\Procedure{Entrofy}{$S, k, \{a_i\}, \{p_i\}, \{w_i\}, \alpha, q$}
\State{$$
    \text{Let }f(X) := \sum_{i} w_i \min{\left(k p_i, \sum_{x\in X} a_i(x)\right)}^\alpha
$$}
\State{Initialize $X \leftarrow \emptyset$}
\While{$|X| < k$}
\State{Let $d(x) := \Delta f(X, x)$ for all $x \in S \setminus X$}
\State{Let $Q := \{x~|~x \in \text{top-}q\text{ quantile of } d(x)\}$}
\State{Select $x$ uniformly at random from $Q$}
\State{$X \leftarrow X \cup \{x\}$}
\EndWhile{}
\State{\textbf{return} $X$, $f(X)$}
\EndProcedure{}
\Statex{}
\Procedure{Entrofy-MC}{$n, S, k, \{a_i\}, \{p_i\}, \{w_i\}, \alpha, q$}
\For{$i \in 1 \dots n$}
\State{$X[i], F[i] \leftarrow \textsc{Entrofy}(S, k, \{a_i\}, \{p_i\}, \{w_i\}, \alpha, q)$}
\EndFor{}
\State{\textbf{return} the $X[i]$ with largest $F[i]$}
\EndProcedure{}
\end{algorithmic}
\end{algorithm}

\subsection*{Encoding attributes}
The algorithm described in the previous section gives a basic framework with which to select candidates given multiple binary attributes.  In this section, we develop extensions of the method to improve its use in practical applications.

\subsubsection*{Non-binary attributes}
The algorithm as described above operates only on binary attributes $a_i$, but many quantities of interest take non-binary values.  Here, we describe methods to convert non-binary attributes into binary values that can be consumed by the algorithm.
We distinguish between two types of non-binary attributes: \emph{categorical} and \emph{ordinal}.

A categorical attribute takes values from a discrete, unordered set of possibilities.
Examples of categorical attributes include a candidate's home institution, area of study, or gender (while we appreciate that gender is not a discrete concept, we advise against modeling gender as a continuous variable here because the proposed algorithmic framework requires an order relation over continuous values).
Categorical attributes can be readily converted into binary attributes by applying a \emph{one-hot encoding}, effectively translating a single variable with some $m$ possible outcomes to $m$ (mutually exclusive) variables each with 2 possible outcomes.

An \emph{ordinal} attribute takes values from a potentially infinite but ordered set, e.g., real numbers.
Examples of ordinal attributes include age, publication count, \emph{etc}.
Ordinal attributes can be binarized by first quantizing the observed values down to a finite, categorical set, and then applying the previously mentioned one-hot encoding scheme.
Quantization thresholds can have a significant impact on the behavior of the algorithm.
In the absence of prior knowledge supplied by the user, we partition the space between the observed minimum and maximum values into $m$ bins of equal length.
When non-binary attributes are binarized, the same weighting coefficient $w_i$ is applied to all corresponding terms in the objective function.

Note that small changes to the boundary positions of histogram bins can produce large changes in the binary encoding and resulting solution.
We advise users to be cautious when dealing with ordinal data.
In our experience, ordinal values are relatively uncommon, and do not present substantial difficulties in practice.

\subsubsection*{Correlations between Attributes}

In practice, one might be interested in attributes that are correlated in some way. This could involve intrinsic correlations in the input data, but could also be a design consideration. For example, one might wish to ensure that the output set contains both junior and senior attendees who identify as women. While the algorithm itself cannot take into account correlations between attributes, this is easily solved during engineering of the attributes. For two attributes $i$ and $j$ with $N_i$ and $N_j$ distinct possible values per attribute, respectively, one can combine both attributes during binarization into a single new attribute with $N_i N_j$ possible values. Instead of setting targets on each individual attribute separately, one may then set targets on the possible values for the combined attribute, which now includes all possible combinations between $i$ and $j$.  

\section*{Results}
\label{sec:results}
\subsubsection*{Experiment 1: Simulating A Single Solution with Varying Noise}
\label{sec:simulations}

We performed several controlled experiments with simulated data sets to demonstrate that our algorithm can successfully recover a solution, i.e., an optimal set of participants embedded in a larger set of random data. We simulated a hypothetical data set of candidates with two categories, each of which comprises two attributes (denoted ``yes'' and ``no'').

We first simulated a single optimal solution to be planted in the data set based on a set of target fractions for each category, to ensure that an optimal solution could in principle be found in every simulation. The objective of these simulations was not to present realistic circumstances (where target fractions are far from the properties of the input set and an optimal solution might not exist), but to explore situations where the algorithm fails when in principle, it should be able to find a solution. To each simulation, we then added a number of candidates with attributes that are randomly selected based on some (other) set of input fractions. These additional candidates effectively act as a type of noise. In realistic situations, ensembles of candidates might be imbalanced with respect to the target fractions; indeed, it might be a stated goal of the selection procedure to address imbalances. Hence we perform simulations with data added to the solution with properties both similar and very dissimilar to the target fractions.
In addition, we varied both the number of participants in the solution set as well as the number of participants added as ``noise''.

\subsubsection*{Set-up}
We vary target fractions between $0.1$ and $0.5$ (because our categories are binary, they are symmetric about $0.5$). We also vary fractions of the attributes of additional random samples added to the solution between $0.0$ and $0.5$, since it is in principle possible that members with certain attributes exist only in the solution and not elsewhere in the set of candidates. We vary the number of participants in the output set (i.e., the solution) between $10$ and $100$, since we deem it unlikely that the algorithm will be used for significantly larger target sets. We let the number of random samples added to the solution vary between $1$ and $1000$. We also varied the parameter $\alpha$ used in the objective function between $0.1$ and $1.0$. For each combination of parameters, we perform $100$ simulations, leading to a total of $24.3 \times 10^{6}$ simulations. 
For each simulation, we run \entrofy\ exactly once. We then computed the value of the objective function for the solution embedded in the data, and calculated the difference between the solution found by \entrofy\ for this data set and the objective on the solution we embedded. When this difference exceeds $0$, we deemed the solution found by \entrofy\ to be a failure.
When the difference is 0, this indicates that \entrofy\ found either the planted solution set, or one of equivalent objective value, which we take as success.

\subsubsection*{Results}
\begin{figure}[htbp]
\begin{center}
\includegraphics[width=3.5in]{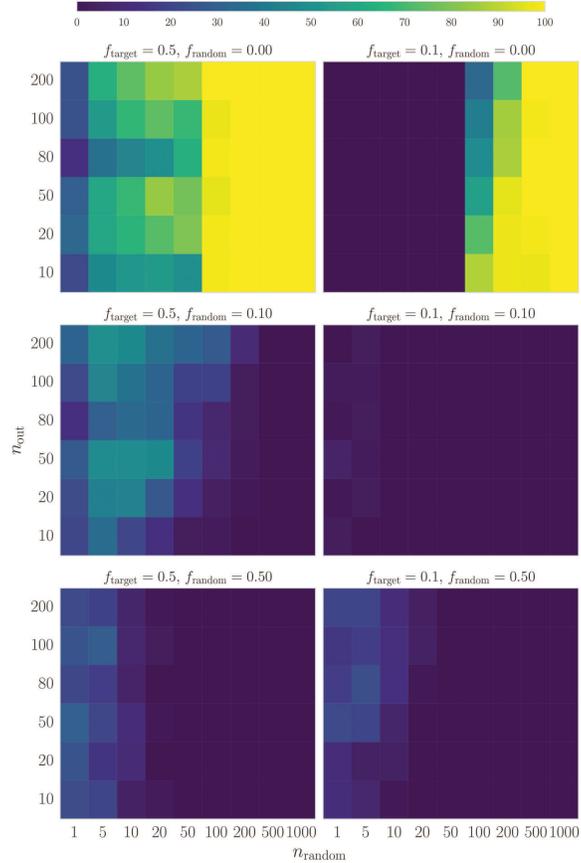}
\caption{Failure rates for simulations with different parameters. On the x-axis, we plot the number of random samples added to each solution $n_\mathrm{random}$ to simulate noise in the data. On the y-axis, we show the number of participants in the output set $n_\mathrm{out}$ (equivalent to the size of the solution embedded in the whole data set). The size of the data set for each simulation is the sum of $n_\mathrm{out}$ and $n_\mathrm{random}$. For all simulations, the target fractions of the output set as well as the random fractions of one category are kept constant at $0$, while values of both target and random fractions are varied for the other category. The left-hand column shows results where the target fraction is kept constant at $0.5$, while the random fraction is varied between $0$ and $0.5$. In the left-hand column, we show similar plots for a target fraction of $0.1$ instead.} 
\label{fig:experiments1}
\end{center}
\end{figure}

In Fig \ref{fig:experiments1}, we present the results of our simulations. Here, we kept the target fractions as well as the random fractions (defining the properties of the randomly added candidates) of one category constant at $0.5$ and varied the target and random fractions of the other attribute to explore the effect both have on the solution. We find that the strongest failure mode occurs when the fraction of the candidates added to the solution for this attribute is $0.0$. In this case, the solution deliberately embedded in the data is the only set of participants yielding an acceptable value of the objective function. Because we use quantiles for tie-breaking when candidates have very similar attributes, this makes finding the optimal solution near-impossible when the set of candidates lacking the attribute in question is very large. Here, the algorithm will almost always fail on a single try, even if the target fraction of that attribute is as low as $0.1$. There appears a fairly sharp phase transition when $100$ or more random samples are added to the output set, no matter the size of the output set. For random ensembles larger than that, the algorithm will fail nearly $100\%$ of the time. However, we caution the reader that the exact position of this phase transition is not well determined, since the grid used in exploring parameter is fairly coarse (and uses steps of either $0.1$ or $10$ for almost all parameters). 

When the random fraction is not $0$, the rate of failure is lowest when it matches the target fractions, and highest when random and target fractions are most dissimilar. In all cases, failure rates are $<50\%$ on a single run of \entrofy\ on a given data set, and failure rates are higher when the size of the random sample added to the solution is small compared to the size of the target output set. This, too, is related to the number of equally optimal solutions in the data set: when the size of the random sample added to the solution is large, and random and target fractions are not too dissimilar, there will be a number of possible solutions, and the probability of the algorithm finding one of them is comparatively large. Conversely, when there are only few samples added to the solution embedded in the data, that solution is likely the only acceptable participant set for \entrofy\ to find, and it will be less probable that the algorithm finds exactly that solution. As expected, this effect is exacerbated if the properties of the random sample and the target fractions differ strongly.

The behaviour described above is largely independent of the parameter $\alpha$. For $\alpha$ between $0.1$ and $0.5$, the results from the simulations are largely consistent. Only when $\alpha = 1$, the failure rate increases significantly, overall by about $5$ failures for each combination of parameters. This is expected, since $\alpha$ changes the shape of the objective function in a way that will make it slightly harder for the algorithm to find the optimal solution when $\alpha = 1$. Thus, we recommend $\alpha = 0.5$ as a reasonable value reliably returning optimal solutions.

\subsection*{Experiment 2: Simulating Multiple Runs}
In practice, running \entrofy\ several times on the same data set and choosing the solution with the maximum value of the objective function is a simple way to mitigate failures on single runs. While this increases runtime, the additional computational cost is small enough on all reasonable data sets (up to at least $1500$ candidates in the input set) to merit the increase in accuracy.

\subsubsection*{Set-up}
In order to test how many simulations are generally necessary to make success highly probable, we picked a case where nearly half of the simulations failed and tested for success as a function of the 
number of times \entrofy\ is run on a single data set, $n_\mathrm{trials}$. We chose a case with an output set of $100$ participants and target fractions of $0.5$ for each attribute in both categories. Again, we simulated a solution with these parameters, and then added $5$ additional candidates to the set distributed with one attribute split $0.1$ (``yes'') and $0.9$ (``no''), the other $0.5$ for ``yes'' and ``no'' both. In our original simulations with $\alpha = 0.5$, this lead to $49$ failures, the highest rate of failure in simulations with this particular combination of target and random fractions. As described above, we compute the objective function for each embedded solution, and subsequently compare with the score returned by \entrofy. In this experiment, however, we vary the number of trials used in \entrofy\ for computing the objective score, between $1$ and $200$, and run $100$ simulations for each value of $n_\mathrm{trials}$. 

\subsubsection*{Results}

In Fig \ref{fig:experiments1}, we present the results of these simulations. Consistent with our previous results, nearly half of the \entrofy\ runs fail to find the optimal solution when the algorithm is run only once. As soon as multiple trials are used, however, the failure rate drops sharply and reaches $0$ at around $10$ iterations. Thus, in practice, allowing \entrofy\ to run $\sim 10$ times and report the best out of those runs allowed us to successfully find the embedded optimal solution in all test cases. In general, this value might depend on the number of distractors, the number of categories and the overall properties of the input set compared to the targets, thus in practice very complex selection procedures might benefit from running additional trials.

\begin{figure}[htbp]
\begin{center}
\includegraphics[width=5in]{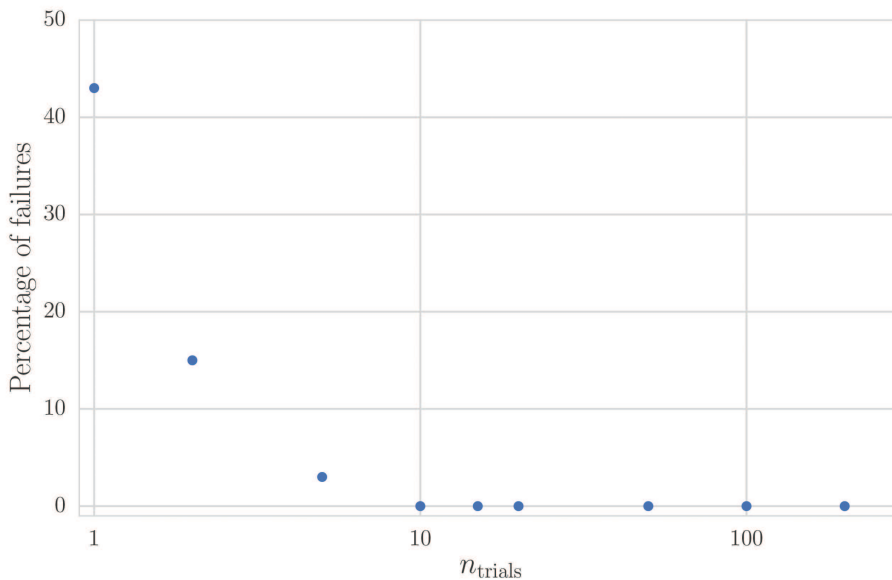}
\caption{The percentage of failed \entrofy\ runs versus the number of 
trials used to find the optimal solution. This simulation uses an optimal solution of $100$ participants in a set of $105$ candidates, where the target fractions of $0.5$ for the defining attribute are dissimilar from those of the set of candidates added to the solution (input fraction $0.1$). We choose $\alpha = 0.5$ and compute the percentage of failures as a function of the number of \textit{Entrofy} runs $n_\mathrm{trials}$ on each simulated data set, using $100$ simulations for each value of $n_\mathrm{trials}$.
For $10$ or more trial runs, \entrofy\ almost always finds the optimal solution.}
\label{fig:experiments2}
\end{center}
\end{figure}

\subsection*{Case Study: Astro Hack Week}
\label{sec:casestudy}

Astro Hack Week is a five-day workshop started in 2014 that recurs annually, with four completed workshops so far. It focuses on data-intensive research in astronomy and has multiple goals: education of the community in the most recent data analysis methods, communicating and encouraging best practices for research, fostering active collaborations and peer learning within astronomy and between astronomy and adjacent disciplines, and networking (for more details, see~\cite{hackweekpaper}). Because of its highly interactive nature and its over-subscription by a factor of $>2$, cohort selection is a central problem during conference organization. For the events in 2016 and 2017, we adopted the approach advocated in this paper: a staggered procedure of quality control followed by optimization based on the goals of the conference. Applicants were informed about the purpose and use of the data being gathered, and filled out a questionnaire including questions probing their motivation and goals for Astro Hack Week, as well as questions relating to their data science-related skills and (optionally) their demographic background.

In the first step, we performed a blind selection of candidates (e.g. their names and institutional affiliations were unavailable to reviewers) based on their responses to several questions about their motivation and goals for their attendance to Astro Hack Week. This served largely to remove spam responses in our openly available online registration form, and remove candidates whose goals diverged strongly from that of the workshop. In total, we selected all 110 candidates in 2016 (155 out of 160 candidates in 2017) in this step, since the workshop has few prerequisites and rejections were only performed for duplicate applications and the rare spam application.

The nature of the workshop requires maximizing diversity over a range of different axes, including demographic diversity (of both race and gender in both years, as well as geographic location in 2017), knowledge of relevant data analysis methods (machine learning, statistics and programming), academic seniority (including all academic ranks from undergraduate students to senior faculty, as well as non-academic roles), and previous attendance at this or a similar event. In total, we optimized over 8 categories (2017: 9 categories), with between 2 and 6 possible options each. 
We pre-selected 11 candidates (2017: 9 candidates) including the scientific organizing committee and selected the remaining cohort of 38 participants (2017: 45 participants) using the algorithm defined in \nameref{sec:algorithm}. In order to quantify similarity between our targets and the distribution of candidates in both our input set as well as our participant set, we define a distance between a candidate set and the targets as 
\begin{equation}
d(X) = \frac{1}{N}\sum_{i=1}^{N} \frac{1}{n_i}\sum_{j=1}^{n_i}\left| \sum_{x \in X} \frac{a_{i,j}(x)}{|X|} - p_{i,j} \right| 
\end{equation}
for $N$ categories with $n_i$ possible attributes each. If $X$ denotes the output set, then this quantity measures the average distance between the targets and the set of selected candidates and is ideally $d(X)=0$. The same quantity can be computed for $X = S$, i.e., the deviation of the entire set of candidates from the targets before selection.

\begin{figure*}[h!]
\begin{center}
\includegraphics[width=15cm]{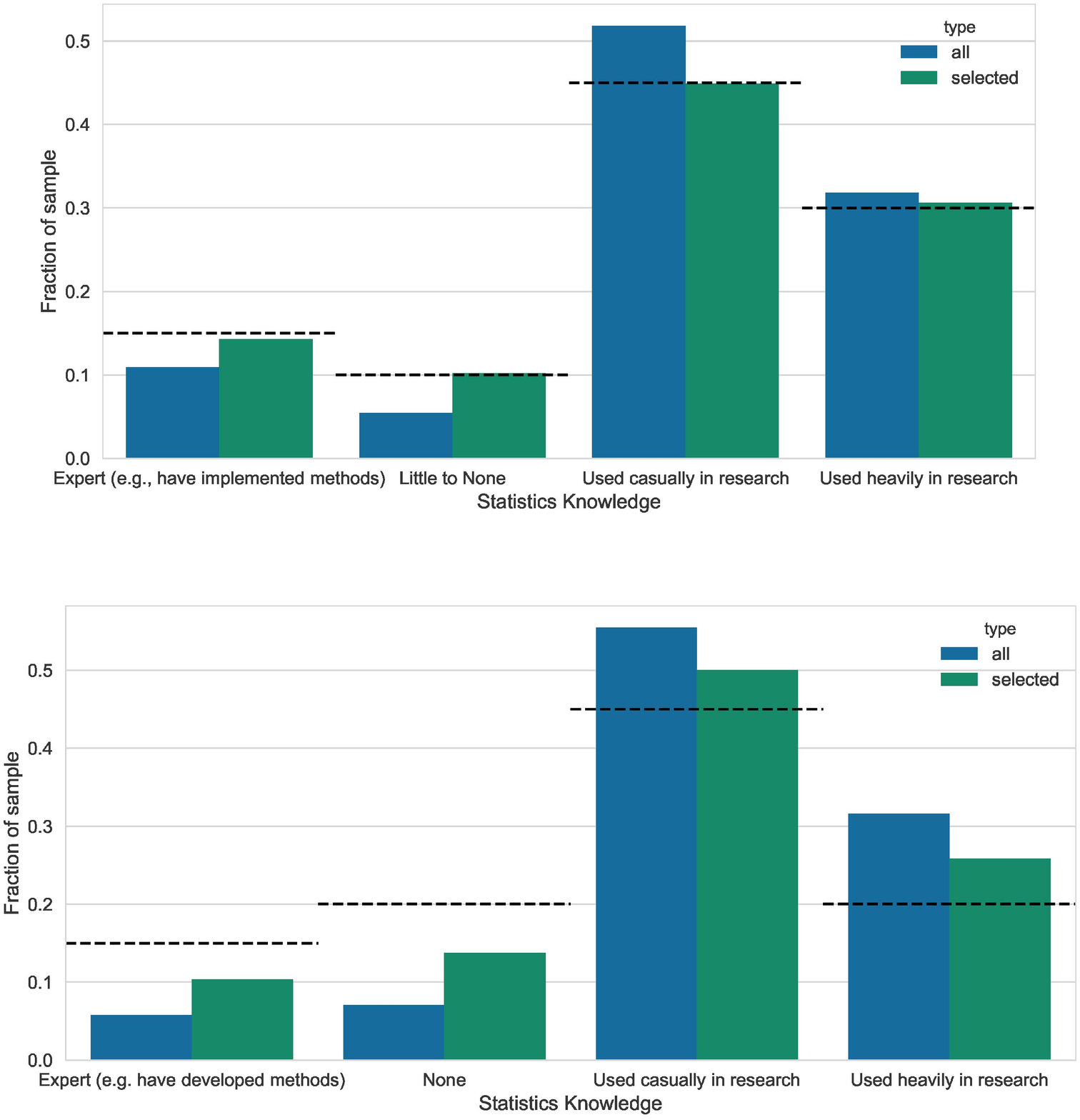}
\caption{This figure shows responses to a question from the registration form asking candidates to self-report their statistics knowledge before the workshop, for both the 2016 workshop (upper panel) and the 2017 workshop (lower panel). In blue, we plot the mean and standard deviation of 1000 trials, where the selection was performed randomly.%In blue, we plot the fraction of candidates for each answer in the candidate set before performing automated selection. 
For comparison, the fraction of answers from the selected cohort is shown in green, and the targets set by the organizing committee based on the workshop goals is plotted as a dashed black line.
In both cases, the selected cohort is notably closer to the pre-defined targets than the original candidate set, though the effect is much stronger in the 2016 case.
The distance between the set of selected candidates and targets is $d(X)=0.016$, compared with $d(S)=0.173$ for the full set of candidates for the 2016 workshop, whereas the distance between selected candidates and targets is $d(X)=0.22$ for the 2017 workshop, with an original distance between targets and input set of $d(S)=0.44$.}
\label{fig:entrofy_stats}
\end{center}
\end{figure*}

In Fig \ref{fig:entrofy_stats}, we present the results of the selection procedure using one particular category as example.
We find that for both workshops, the set of acceptable candidates (or a random selection thereof) diverges significantly from the ideal cohort as defined by our workshop goals.
For this particular category, we find an initial distance between the set of candidates and the targets of $d(S)=0.043$ in 2016, and $d(S)=0.11$ in 2017.
In the set of selected candidates, the distance is reduced by more than a factor of 10 to $d(X)=0.0041$ (2016), but only by about a factor of 2 to $d(X)=0.054$ in 2017.
The latter is due to the fact that despite the larger candidate set in 2017, the overall divergence between input set and targets was much larger, to a point where it became impossible to find a solution close to the targets.

Repeating this analysis using all categories yields distances $d(S)=0.074$ in 2016 and $d(S)=0.148$ in 2017.
Note, however, that the 2017 data set includes an additional category, geographic location, for which we had set ambitious targets not supported by the set of candidates.
Even without inclusion of that category, the average distance between input set and targets is still $d(S)=0.12$, almost twice as high as for the 2016 set, indicating that perhaps the 2017 workshop attracted a slightly different population from the earlier workshop.
Part of this might be explained by the much higher fraction of graduate students who applied in 2017 (62\% compared to 50\% in 2016) and the lower target for that particular category set (0.3 in 2017 versus 0.4 in 2016).
    
Overall, Entrofy found a participant set in 2016 that reduced the distance between participants and targets to $d(X)=0.028$ in the output set. While there seems to be no optimal set of participants in the data, the application of \entrofy\ has resulted in a set of participants that are a considerably closer fit to the desired cohort targets than we would have gotten if we selected a cohort that matched the properties of the applicant pool.
In 2017, conversely, the overall change in distance is much smaller, by less than a factor of 2 to $d(X)=0.10$.
The differences can be explained by the constrained nature of the input data set: with only an over-subscription of a factor of $\sim 2$ and for some categories a large divergence between input set and targets, an optimal cohort may simply not exist.
Additionally, differences may result from trying to globally optimize all eight categories at the same time.

The concave transformation weights participants higher if they improve a category still far from the target, sometimes at the expense of introducing small deviations in other categories.
Note also, that selected candidates declining attendance necessitate re-selection (also performed with Entrofy).
If candidates preferentially drop out in certain categories that are already underrepresented, this may exacerbate the distance between participant set and pre-defined targets.

\section*{Discussion}
\label{sec:discussion}

The cohort selection procedure we have described can be effectively applied to produce diverse selections from a pool of qualified candidates.
In this section, we briefly overview some of the beneficial side-effects of the proposed method, describe caveats and recommendations for effective use in practice, and highlight directions for future work.

%% entrofy is a building block for tie-breaking
%% can help clarify selection decisions
%% continued recap
\subsection*{Benefits of algorithmic cohort selection}
There is now a wealth of available research, especially from the employment literature, that recommends making assessment and decision processes as structured and quantitative as possible to facilitate better decision outcomes ~\cite{sunstein2015wiser}.
This allows committees to be accountable to themselves and to the candidates they are assessing.
Because goals and requirements may differ for different scenarios, experimentation with strategies, questions, weights and targets should be encouraged, as should be critical evaluation both during and after the selection.
This is especially true for recurring selection events (e.g., annual conferences), where cohorts can be compared laterally.
%% entrofy makes decisions explicit
Entrofy fits within this framework in three ways, providing benefits to selection committees beyond the output of the algorithm.

First, it helps committees develop the data set and the language around decision processes to enable evaluation and discussion.
Because Entrofy requires committees to explicitly attach numerical targets and weights to attributes, discussions shift toward the goals and requirements of the conference, workshop, or degree program. In particular, committees discuss how to translate their aims into questions asked of candidates and to establishing targets or weights for Entrofy.
Drawing on Astro Hack Week as an example, participant questions about computer programming skill have shifted from being broad and undirected (e.g., \emph{how many years of programming experience?}) to more fine-grained, objective, and well-calibrated questions that people can realistically self-report (e.g., whether they have written certain programming constructs like functions or classes).
Similarly, the procedure encouraged the the committee to critically think about the results of the selection, and allowed it to first visualize cases of interest (e.g., whether there were correlations between career stage and gender) and subsequently discuss the inclusion of a joint category in the selection in order to ensure representation of gender minority participants among all career stages.
Detailed evaluations and discussion about workshop goals, in turn, have helped clarify the selection process, but unexpectedly aided the committee in other crucial tasks like program decisions. 
Overall, it has provided a framework for critical post-hoc evaluation, comparing selection procedures and workshop outcomes over successive years to improve the process or experiment with alternative strategies.

%% useful side-effects, esp for repeated events
A second positive side-effect is that the associated data collection and visualization can help the committee identify potential biases inherent in applicant pool and selection procedure.
With respect to Astro Hack Week, the committee used data visualizations made as part of the Entrofy process to identify a lack of candidates from underrepresented racial and ethnic minorities.
This in turn led to targeted outreach efforts toward those groups, as well as directed fund-raising efforts to aid students from disadvantaged backgrounds and smaller institutions.  

%% transparency and accountability
Third, it provides algorithmic transparency and accountability for the selection procedure. In the case study of Astro Hack Week during which participants were told that Entrofy had been part of the selection procedure, an online survey was administered at the end of the workshop to gauge participants' 

82 percent agreed with the statement "I think that using an algorithm for selection makes the process more transparent" (see Table \ref{tab:outcomes}). Similarly, 81 percent reported agreement with the statement, "I believe the selection for AstroHackWeek was fair". Because academic meetings are an important venue for discussion, collaboration, and career development, both conference organizers and research communities have a vested interest in making conferences and workshops equitable to candidates from all backgrounds and at all stages in their careers. An overwhelming 97 percent of respondents in our case study agreed with the statement, "I think Astro Hack Week benefitted from a wide range of backgrounds and experiences of the attendees."
The underlying justification for using a two-stage selection process is likely to apply to many other situations where decision makers must select a subset of candidates, including hiring, speaker selection for seminar series, and degree program admissions procedures.
Splitting the selection into two steps and framing the second part as a clearly defined mathematical procedure provides this transparency: if a candidate is rejected, that decision can be traced back to either the merit-based selection stage or the randomized tie-breaking stage (Entrofy).

\begin{sidewaystable}
\renewcommand{\arraystretch}{1.3}
\footnotesize
\caption{}
\begin{threeparttable} 
\begin{tabularx}{22cm}{p{2cm}p{3.5cm}p{3.5cm}p{0.5cm}p{3.5cm}p{3.5cm}p{3.5cm}}
\toprule
%\bf{Parameter} & \bf{Meaning} & \bf{Probability Distribution} \\ \midrule
 & \multicolumn{3}{c}{Negative outcomes of using Entrofy?} & \multicolumn{3}{c}{Positive outcomes of using Entrofy?} \\
 \midrule
 & \textit{I think the wide range of backgrounds made the meeting too unfocused.} & \textit{I was more uncomfortable because attendees came from a range of backgrounds.} && \textit{I think AstroHackWeek benefitted from a wide range of backgrounds and experiences of attendees.} & \textit{I think that using an algorithm for selection makes the process more transparent.} & \textit{I believe the selection procedure for AstroHackWeek was fair.} \\
 \midrule
 Strongly Disagree & $14$ ($42\%$) & 0 (0\%) && 12 (36\%) & 0 (0\%) & 1 (3\%) \\
 Disagree & 9 (27\%) & 0 (0\%) && 6 (19\%) & 0 (0\%) & 0 (0\%) \\
 Somewhat Disagree & 5 (15\%) & 1 (3\%) && 3 (10\%) & 3 (10\%) & 0 (0\%) \\
 Somewhat Agree & 4 (12\%) & 4 (12\%) && 5 (15\%) & 4 (13\%) & 2 (6\%) \\
 Agree & 0 (0\%) & 10 (30\%) && 4 (12\%) & 9 (27\%) & 11 (33\%) \\
 Strongly Agree & 1 (3\%) & 18 (54\%) && 3 (9\%) & 14 (42 \%) & 14 (42\%) \\
 Don't Know & NA & NA && NA & 3 (10\%) & 5 (15\%) \\
\bottomrule
\end{tabularx}
   \begin{tablenotes}
      \item{Respondents to a post-event evaluation survey indicate their level of agreement with the statements in the header. $N=33$; response rate $66\%$}
\end{tablenotes}

\end{threeparttable}
\label{tab:outcomes}
\end{sidewaystable}

\subsection*{Recommendations for effective use of Entrofy}
While controlling for human bias is one of the primary goals of implementing the algorithm developed here, the algorithm itself is not a complete solution to the problem.
Careful human judgment is required at several points in the procedure, most notably in the initial selection for merit, the selection of categories to be included in the algorithm, and establishing the size of target proportions.
Even for blinded reviews in the first stage, assessors must be vigilant in checking their biases or the set of acceptable candidates may be distorted in undesirable ways.
It is worth noting explicitly that if the initial recruitment and/or selection are strongly biased, Entrofy is unlikely to be able to find a solution close to the targets, even if those targets are deemed fair. 
In other words, in cases where gender parity is desired, but the recruitment and/or blind review process are unable to attract a sufficient number of qualified women, Entrofy will not be able to deliver a cohort with gender parity.
When the initial selection for merit prefers participants with certain characteristics at the expense of others, it will likely make finding an optimal solution (or a solution close to optimal) impossible.
Given the targets, Entrofy may be able to correct biases introduced during the first step to a certain degree, but that is \textit{not} its intended purpose.
However, categories where the targets are far away from the input set can be easily diagnosed from plots like shown in Fig~\ref{fig:entrofy_stats}, which can be generated by the \entrofy\ software package.
In these cases, we suggest that organizers critically examine their initial recruitment and selection procedure for biases that could have resulted in generating a poorly representative set of meritorious candidates. For more practical advice for the application of Entrofy in realistic problems, see also the Supplementary Materials.

%% be responsible in defining attributes
Similarly, it is the selection committee's responsibility to define categories and targets that do not favour one group over another in ways that are misaligned with the goals of the selection. For example, setting a minimum threshold on a particular skill or knowledge may systematically disadvantage certain categories of candidates if these candidates have had less access to opportunities to learn that skill or knowledge due to structural inequalities. 
Overall, it is likely that this procedure is most useful to committees already sensitized to issues around diversity and committed to improving their procedures.
In this case, objective gains made by employing Entrofy as part of the overall strategy may be limited, but as discussed above, usage of the software and algorithm may free up the committee's mental capacities for critically evaluating selection procedures.

%% continuing on defining attributes
As a matter of ethics, using a blind review followed by an unblinded algorithmic selection based on desired characteristics could have either a beneficial or \emph{detrimental} outcome for individual participants and the cohort taken collectively.
The target characteristics are determined by the selection committee, which leaves them subject to that group's judgment.
To the extent that Entrofy is being used as envisioned, it is likely to lead to increased fairness for individual applicants around known biases and a more desirable cohort experience for those selected.
However, in certain situations, blind review has been found to make selected groups \textit{decrease} diversity~\cite{behaghel2015unintended}.
This is especially the case for selection committees who already have a strong commitment to diversity and mechanisms in place to test for biases.
In these cases, a blinded first step may not be the optimal approach.
We note that using demographic criteria like race, gender, and other identity statuses protected under employment law as targets in hiring decisions may require legal counsel beyond the scope of this software review.

Note, however, the larger and more biased the preselected cohort is compared with the targets, the more likely it may be that finding an optimal subset becomes impossible. 

%\subsubsection*{OTHERS???}

\subsection*{Future work}
%% future work: assessing long-term outcomes
While we have shown here that the algorithm correctly selects cohorts that closely adhere to the chosen targets where possible, the long-term effects of the proposed method on the resulting cohorts and its participants have not been systematically studied.
In general, assessing the outcomes of a selection procedure is difficult for two reasons.
First, defining ``success'' is non-trivial in many selection problems.
Second, control groups almost never exist: we simply do not know how the other job candidate would have performed, or how successful a conference would have been if a different set of speakers had been selected, or a different selection procedure was employed.
Even clear evidence that conference attendance has a positive effect on early-career researchers is scarce.

In the future, the algorithm proposed here might open this line of inquiry in new ways.
In particular, a crucial problem in studying conference outcomes is the selection problem: have participants been selected because they were already successful or has the conference improved their chances of success?
The two-step procedure advocated here yields a pool of roughly equally qualified candidates some of whom were rejected (and did not attend) and others who were selected for attendance based on criteria besides merit (most of whom attended).
Following the outcomes of individuals in both groups may provide a relatively clean data set for studying the effect conference/workshop/degree program attendance has on measures of success such as citations and career objectives. 
Anecdotally, we know from discussions and open-ended survey questions that participants favour diverse workshops.
In particular, early-career researchers who identify as a gender and/or racial/ethnic minorities remarked positively on the representation of diverse demographic groups at Astro Hack Week.
However, it is currently too early to assess the impact Astro Hack Week might have on the participants' careers. 

%% more future-work: comparing directly to manual selections
Much of the discussion about the way \textit{selection committees} engage with Entrofy and how it changes committee members' thinking about the procedure  either draws on results from the hiring literature or is anecdotal.
Similarly, we have not conclusively shown that a committee using Entrofy makes objectively better decisions about a cohort than a committee performing a more traditional selection.  
In particular, it would be instructive to see a committee that traditionally has reported successes in selecting a diverse set of (unblinded) candidates adopt the procedure and compare the results.
It would also be rewarding to study how the decision making process changes when committees utilize the approach suggested here, and if it indeed leads committees to ask critical questions about their selection, rephrase part of it in quantitative ways and use the data to improve future workshops.

%% more future work: expand beyond workshops to other cohort selection problems
Finally, cohort selection is a problem in many different contexts beyond participant selection for scientific workshops, which originally motivated this line of research.
In the context of academic selections, the proposed framework can be extended to traditional conferences where talks, posters and papers are actively filtered.
The similarities between this application and workshop participant selection invite a closely analogous procedure.

%% continued: other cohort selections, admissions
More generally, we envision that any cohort selection process could be a potential use case.
In particular, we would like to explore applications of Entrofy to undergraduate or graduate school admissions in which candidates are generally selected by committees similar in structure to conference selection committees.
In this context, blinding during the first stage is difficult due to the nature of Curriculum Vitae and reference letters. Further, as discussed above, it may produce adverse results in situations where panels are particularly sensitive to diversity already.
There is still merit, however, in making the process more quantitative.
The overall applications could still be scored, and that scoring interrogated.
As in the context of hiring, there is likely to be an large intrinsic variance in scores, and in general ranking decisions by committees may have low predictability especially when the decision process includes unstructured interviews, as e.g.\ a natural experiment at the University of Texas Medical School has shown~\cite{devaul1987medical}.
Especially in highly competitive settings, which also includes for example grant proposals or fellowship applications, there is likely a large pool of highly qualified candidates and selection procedures might benefit considerably from accounting for committee variance in initial merit decisions by setting a fairly tolerant cut-off for acceptable candidates and letting Entrofy decide between them based on other selection goals.

%% more cohort selection problems: employment
We do not anticipate Entrofy to be useful for selecting employees in a classical hiring context, because it relies on maximizing a \textit{cohort} over multiple categories, whereas hiring decisions are usually performed for individual positions.
However, Entrofy could potentially be used in producing a balanced shortlist of candidates to be invited to interview or in cohort hiring scenarios as with first year law firm, banking, and consulting candidates ~\cite{rivera2015}.

\section*{Conclusions}

Cohort selection is and remains an intrinsically human, biased, and difficult problem.
A plethora of evidence suggests that judgments based on the assessor's experience and intuition often lead to selections that align with the assessor's biases and lack predictive power with respect to selection outcomes.
We suggest here that a two-step procedure based on a blind merit selection followed by an algorithmic cohort selection based on extrinsic criteria can produce cohorts whose attributes align with uniquely defined cohort characteristics.
We have presented a new algorithm and software, Entrofy, that automates the second part of this process, have shown that it has found optimal solutions in practice, and have argued that this process provides accountability, transparency, and empowers humans to overcome implicit biases in selection processes.
The selected cohorts match pre-defined targets in both simulations and our case study, insofar as an optimal cohort exists.
We propose that our solution allows for improved control of human biases during cohort selection, as well as greater accountability and fairness for candidates.
However, research in academic cohort selection remains scarce and should be a priority for the future.

\section*{Acknowledgments}
The work presented here grew directly from a collaboration that started at the Astro Hack Week 2015. %\textbf{We thank the anonymous referees for their thoughtful and very useful comments on the draft manuscript.} 
We are indebted to these researchers for their constructive feedback on the algorithm and software as well as many useful discussions: Kelle Cruz, Katalin Takats, R. Stuart Geiger, Alexandra Paxton, Josh Greenberg, Tal Yarkoni, Josh Bloom, David W. Hogg, Phil Marshall, Jake Vanderplas, Lucianne Walkowitz, Adam Miller, Erik Tollerud, Erik Jeschke and Ralph Wijers. \\
\textbf{Funding:} DH, BM and LN acknowledge support by the Moore-Sloan Data Science Environment at NYU. DH was partially funded by the James Arthur Postdoctoral Fellowship at NYU. DH acknowledges support from the DIRAC Institute in the Department of Astronomy at the University of Washington. The DIRAC Institute is supported through generous gifts from the Charles and Lisa Simonyi Fund for Arts and Sciences, and the Washington Research Foundation\\
%\textbf{Author Contributions} DH led the project and the writing, performed the experiments and the data analysis for both simulations and case study. BM developed the core algorithm and wrote the methods section. DH and BM co-wrote the associated software. LN provided social science context in the writing and obtained IRB permission for review of Astro Hack Week data. \\
%\textbf{Competing Interests} The authors declare that they have no competing interests.\\
\textbf{Data and materials availability:} All data needed to evaluate the conclusions in the paper are present in the paper and/or the Supplementary Materials. The software tool accompanying this paper, along with additional simulations and all underlying code, is available at \url{https://github.com/dhuppenkothen/entrofy}. Additional data available from authors upon request. The full selection procedure for Astro Hack Week 2016 has been described in a blog post (\url{https://danielahuppenkothen.wordpress.com/2016/11/16/workshop-participant-selection-from-start-to-finish-an-example/}). Additionally, fully reproducible code exists for another conference~\url{https://github.com/dhuppenkothen/PyAstro17ParticipantSelection}.

\nolinenumbers

% Either type in your references using
% \begin{thebibliography}{}
% \bibitem{}
% Text
% \end{thebibliography}
%
% or
%
% Compile your BiBTeX database using our plos2015.bst
% style file and paste the contents of your .bbl file
% here. See http://journals.plos.org/plosone/s/latex for 
% step-by-step instructions.
% 

%\input extra_refs
\bibliography{refs}

%\begin{thebibliography}{10}

%\bibitem{bib1}
%Conant GC, Wolfe KH.
%\newblock {{T}urning a hobby into a job: how duplicated genes find new
%  functions}.
%\newblock Nat Rev Genet. 2008 Dec;9(12):938--950.

%\bibitem{bib2}
%Ohno S.
%\newblock Evolution by gene duplication.
%\newblock London: George Alien \& Unwin Ltd. Berlin, Heidelberg and New York:
%  Springer-Verlag.; 1970.

%\bibitem{bib3}
%Magwire MM, Bayer F, Webster CL, Cao C, Jiggins FM.
%\newblock {{S}uccessive increases in the resistance of {D}rosophila to viral
%  infection through a transposon insertion followed by a {D}uplication}.
%\newblock PLoS Genet. 2011 Oct;7(10):e1002337.

%\end{thebibliography}

\end{document}